\documentclass[prd,twocolumn,groupedaddress,showpacs,nofootinbib]{revtex4}
\usepackage{graphicx}
\usepackage{dcolumn}
\usepackage{amssymb}
\usepackage{mathrsfs}
\usepackage{amsmath}
\usepackage{epsfig}
\usepackage[dvips]{color}
\usepackage{hhline}

\begin{document}

\title{A left-right symmetric model with $SU(2)$-triplet fermions}

\author{Pei-Hong Gu}
\email{peihong.gu@mpi-hd.mpg.de}

\affiliation{Max-Planck-Institut f\"{u}r Kernphysik, Saupfercheckweg
1, 69117 Heidelberg, Germany}

\begin{abstract}

We consider an $SU(3)_c^{}\otimes SU(2)_L^{}\otimes
SU(2)_R^{}\otimes U(1)_{B-L}^{}$ left-right symmetric model with
three Higgs scalars including an $SU(2)_L^{}$ doublet, an
$SU(2)_R^{}$ doublet and an $SU(2)_L^{}\otimes SU(2)_R^{}$
bidoublet. In addition to usual $SU(2)$-doublet fermions, our model
contains $SU(2)$-triplet fermions with Majorana masses. The neutral
components of the left-handed triplets can contribute a canonical
seesaw while the neutral components of the right-handed triplets
associated with the right-handed neutrinos can contribute a
double/inverse-type seesaw. Our model can be embedded into an
$SO(10)$ grand unification theory where the triplets belong to the
$\textbf{45}= (\textbf{1},\textbf{3},\textbf{1},0)\oplus
(\textbf{1},\textbf{1},\textbf{3},0)\oplus ... $ representations.

\end{abstract}


\maketitle

In the $SU(3)_c^{}\otimes SU(2)_{L}^{}\otimes SU(2)_{R}^{}\otimes
U(1)_{B-L}^{}$ left-right symmetric models \cite{ps1974}, the
standard model (SM) fermions plus the right-handed neutrinos are
placed in the $SU(2)_L^{}$ or $SU(2)_R^{}$ doublets. With an
appropriate Higgs content, we can generate the fermion masses by
supplementing additional fermions or not. The minimal Higgs content
seems to be an $SU(2)_R^{}$ doublet and an $ SU(2)_L^{}$ doublet,
which drive the left-right and electroweak symmetry breaking,
respectively. With the $SU(2)$-doublet Higgs scalars, one can
consider the universal \cite{berezhiani1983} seesaw
\cite{minkowski1977} scenario to generate the SM fermion masses and
the neutrino masses by introducing $SU(2)$-singlet fermions. In the
most popular version of the left-right symmetric models, the fermion
fields only contain the usual $SU(2)$ doublets while the Higgs
fields contain two $SU(2)$ triplets and one $SU(2)_{L}^{}\otimes
SU(2)_{R}^{}$ bidoublet. The right-handed Higgs triplet and the
Higgs bidoublet are responsible for the left-right and electroweak
symmetry breaking, respectively. The charged fermions can obtain
their masses through the usual Yukawa interactions while the
neutrinos can obtain their masses through the type-I
\cite{minkowski1977} and II \cite{mw1980} seesaw mechanisms. In the
original left-right symmetric model, the Higgs content contains two
$SU(2)$ doublets and one $SU(2)_{L}^{}\otimes SU(2)_{R}^{}$
bidoublet. The right-handed Higgs doublet and the Higgs bidoublet
account for the left-right and electroweak symmetry breaking,
respectively. In this scenario, it is difficult to naturally explain
the smallness of the neutrino masses. To solve this problem, one can
introduce gauge-singlet fermions with Majorana masses to realize the
linear \cite{barr2003} and double/inverse
\cite{mohapatra1986,mv1986} seesaw.

In this paper we shall extend the original left-right symmetric
model by $SU(2)$-triplet fermions to generate the small neutrino
masses in a nature way. This idea has been proposed in an early work
\cite{perez2009}. However, the author explicitly broke the discrete
left-right symmetry to protect the left-handed Higgs doublet from a
nonzero vacuum expectation value (VEV). In consequence, a
double/inverse-type seesaw induced by the neutral components of the
right-handed triplets associated with the right-handed neutrinos
becomes the unique origin of the neutrino masses. We shall study the
general scenario where the left-handed Higgs doublet has a nonzero
VEV. Specifically, the neutrinos can also pick up their masses
through a canonical seesaw induced by the neutral components of the
left-handed triplets. The canonical seesaw even can dominate over
the double/inverse-type seesaw. Furthermore, we will embed our model
into the $SO(10)$ grand unification theory (GUT). In the GUT
context, the $SU(2)$-triplet fermions belong to the $\textbf{45}$
representations. A $\textbf{45}$ multiplet also contain a singlet
fermion under the left-right gauge symmetry. The singlet fermions
can result in a double/inverse-type seesaw and a linear seesaw.

We take the notations as follows:
\begin{eqnarray}
&\chi_L^{}(\textbf{1},\textbf{2},\textbf{1},1)=
\left[\begin{array}{c}\chi_{L}^{+}\\
[2mm]\chi_{L}^{0}\end{array}\right]\,,~~
\chi_R^{}(\textbf{1},\textbf{1},\textbf{2},1)=
\left[\begin{array}{c}\chi_{R}^{+}\\
[2mm]\chi_{R}^{0}\end{array}\right]\,,&
\nonumber\\
[2mm] &\phi(\textbf{1},\textbf{2},\textbf{2},~0)=
\left[\begin{array}{cc}\phi_1^{0}&\phi_2^{+}\\
[2.5mm] \phi_1^{-}&\phi_2^{0}\end{array}\right]&
\end{eqnarray}
are the Higgs scalars, while
\begin{eqnarray}
\begin{array}{rcrc}
q_L^{}(\textbf{3},\textbf{2},\textbf{1},+\frac{1}{3})=&
\left[\begin{array}{c}u_L^{}\\
[2mm]d_L^{}\end{array}\right]\,,&
q_R^{}(\textbf{3},\textbf{1},\textbf{2},+\frac{1}{3})=&
\left[\begin{array}{c}u_R^{}\\
[2mm]d_R^{}\end{array}\right]\,,\\
[8mm] l_L^{}(\textbf{1},\textbf{2},\textbf{1},-1)=&
\left[\begin{array}{c}\nu_L^{}\\
[2mm]e_L^{}\end{array}\right]\,,&
l_R^{}(\textbf{1},\textbf{1},\textbf{2},-1)=&
\left[\begin{array}{c}\nu_R^{}\\
[2mm]e_R^{}\end{array}\right]\,,\\
[8mm] T_L^{}(\textbf{1},\textbf{3},\textbf{1},~0)=&
\left[\begin{array}{c}T_{L}^{1}\\
[2mm]T_{L}^{2}\\
[2mm]T_{L}^{3}\end{array}\right]\,,&
T_R^{}(\textbf{1},\textbf{1},\textbf{3},~0)=&
\left[\begin{array}{c}T_{R}^{1}\\
[2mm]T_{R}^{2}\\
[2mm]T_{R}^{3}\end{array}\right]\end{array}\nonumber
\end{eqnarray}
\vspace{-5mm}
\begin{eqnarray}
\end{eqnarray}
denote the fermions. It is convenient to rewrite the fermion
triplets by the matrix representation,
\begin{eqnarray}
T_{L,R}^{}&=&\tau_1^{} T_{L,R}^{1}+\tau_2^{}
T_{L,R}^{2}+\tau_3^{} T_{L,R}^{3}\nonumber\\
[2mm]
&=&\left[\begin{array}{rr}T^0_{L,R}&\sqrt{2}T^+_{L,R}\\
[2.5mm] \sqrt{2}T^-_{L,R}&-T^0_{L,R}\end{array}\right]
\end{eqnarray}
with
\begin{eqnarray}
T_{L,R}^{\pm}=\frac{T_{L,R}^1\mp
iT_{L,R}^2}{\sqrt{2}}\,,~~T_{L,R}^0=T_{L,R}^3\,.
\end{eqnarray}

The right-handed Higgs doublet $\chi_R^{}$ will develop a VEV
\begin{eqnarray}
\langle\chi_R^{}\rangle= \left[\begin{array}{c}0\\
[2mm]v_{R}^{}\end{array}\right]
\end{eqnarray}
for the left-right symmetry breaking. Subsequently, the Higgs
bidoublet will acquire a VEV
\begin{eqnarray}
\langle\phi\rangle=
\left[\begin{array}{cc}v_1^{}&0\\
[2.5mm] 0&v_2^{}\end{array}\right]
\end{eqnarray}
to break the electroweak symmetry. Because the left-handed Higgs
doublet has the trilinear couplings with the right-handed Higgs
doublet and the Higgs bidoublet, i.e.
\begin{eqnarray}
V&\supset&\mu \chi_L^\dagger \phi \chi_R^{} + \tilde{\mu}
\chi_L^\dagger \tilde{\phi} \chi_R^{} +\textrm{H.c.}\,,
\end{eqnarray}
it will pick up an induced VEV \footnote{The electroweak symmetry
breaking can be driven by the left-handed Higgs doublet. The Higgs
bidoublet then can obtain a VEV at the weak scale through its
trilinear couplings with the left- and right-handed Higgs
doublets.},
\begin{eqnarray}
\langle\chi_L^{}\rangle&=&
\left[\begin{array}{c}0\\
[2mm]v_L^{}\end{array}\right]\quad \textrm{with} \nonumber\\
[2mm] && v_L^{}\simeq\frac{\mu v_R^{} v_2^{}+ \tilde{\mu} v_R^{}
v_1^{}}{M_{\chi_L^0}^2}\sim \frac{\mu v_2^{}+ \tilde{\mu}
v_1^{}}{v_R^{}}\,,
\end{eqnarray}
which is a weak-scale value for $\mu\sim \tilde{\mu}\sim v_R^{}$ or
a tiny value for $\mu\sim \tilde{\mu}\ll v_R^{}$. The allowed Yukawa
couplings should be
\begin{eqnarray}
\mathcal{L}_Y^{}&=&- y_q^{}\bar{q}_L^{}\phi q_R^{}-
\tilde{y}_q^{}\bar{q}_L^{}\tilde{\phi}q_R^{} -y_l^{}\bar{l}_L^{}\phi
l_R^{}- \tilde{y}_l^{}\bar{l}_L^{}\tilde{\phi}
l_R^{}\nonumber\\
[2mm] &&-f_L^{}\bar{l}_L^c i\tau_2^{}T_L^{} \chi_L^{}
-f_R^{}\bar{l}_R^c i\tau_2^{}T_R^{} \chi_R^{} +\textrm{H.c.}\,.
\end{eqnarray}
The fermion triplets further have the Majorana masses as below,
\begin{eqnarray}
\mathcal{L}_M^{}=-\frac{1}{4}M_L^{}\textrm{Tr}(\bar{T}_L^{c} T_L^{})
- \frac{1}{4}M_R^{}\textrm{Tr}(\bar{T}_R^{c} T_R^{})
+\textrm{H.c.}\,.
\end{eqnarray}

Through the Yukawa couplings with the Higgs bidoublet, the charged
fermions can obtain their masses, i.e.
\begin{eqnarray}
\mathcal{L}_m^{}&\supset&
-m_d^{}\bar{d}_L^{}d_R^{}-m_u^{}\bar{u}_L^{}u_R^{}-m_e^{}\bar{e}_L^{}e_R^{}+\textrm{H.c.}\,,
\end{eqnarray}
with \begin{subequations}
\begin{eqnarray}
m_d^{}&=&y_q^{}v_2^{}+\tilde{y}_q^{}v_1^{}\,,\\
[2mm]
m_u^{}&=&y_q^{}v_1^{}+\tilde{y}_q^{}v_2^{}\,,\\
[2mm] m_e^{}&=&y_l^{}v_2^{}+\tilde{y}_l^{}v_1^{}\,.
\end{eqnarray}
\end{subequations}
As for the neutral fermions, including the left- and right-handed
neutrinos as well as the the neutral components of the left- and
right-handed triplets, their mass term is given by
\begin{eqnarray}
\label{neutral} \mathcal{L}_m^{}&\supset&
f_L^{}v_L^{}\bar{\nu}_L^{c}T_L^0-\frac{1}{2}M_L^{}(\bar{T}_L^{0})^c_{}T_L^0
-(y_l^{}v_1^{}+\tilde{y}_l^{}v_2^{})\bar{\nu}_L^{}\nu_R^{}\nonumber\\
[2mm]
&&+f_R^{}v_R^{}\bar{\nu}_R^{c}T_R^0-\frac{1}{2}M_R^{}(\bar{T}_R^{0})^c_{}T_R^0+\textrm{H.c.}\nonumber\\
[2mm]
&=&-\frac{1}{2}\left[\bar{\nu}_L^c\,,\,(\bar{T}_L^0)^c_{}\right]
\left[\begin{array}{cc}0&-f_L^{}v_L^{}\\
[2.mm] -f_L^T&M_L^{}\end{array}\right]
\left[\begin{array}{c}\nu_L^{}\\
[2.mm] T_L^0\end{array}\right]\nonumber\\
[2mm]
&&-\frac{1}{2}\left[\bar{\nu}_L^c\,,\,\bar{\nu}_R^{}\,,\,(\bar{T}_R^0)^c_{}\right]\nonumber\\
[2mm] &&\times
\left[\begin{array}{ccc}0&y_l^{T}v_1^{}+\tilde{y}_l^{T}v_2^{}&0\\
[2.mm]y_l^{} v_1^{}+\tilde{y}_l^{} v_2^{}
&0&f_R^\ast v_R^{}\\
[2.5mm]0&f_R^\dagger v_R^{} &
M_R^\dagger\end{array}\right]\left[\begin{array}{c}\nu_L^{}\\
[2.mm]\nu_R^c\\
 [2.mm] (T_R^0)^c_{}\end{array}\right]\nonumber\\
 [2.mm]
 &&+\textrm{H.c.}\,.
\end{eqnarray}
It is straightforward to see the $2\times 2$ mass matrix can induce
a canonical seesaw formula \cite{minkowski1977},
\begin{eqnarray}
\mathcal{L}_m^{}\supset
-\frac{1}{2}m_\nu^{L}\bar{\nu}_L^{c}\nu_L^{}+\textrm{H.c.}~~
\textrm{with}~~m_\nu^L=-f_L^{}\frac{v_L^2}{M_L^{}}f_L^T\,,
\end{eqnarray}
like that in the type-I \cite{minkowski1977} or III \cite{flhj1989}
seesaw extension of the SM. On the other hand, the $3\times 3$ mass
matrix can induce a double/inverse-type seesaw formula
\cite{mohapatra1986,mv1986},
\begin{eqnarray}
\label{doubleinverse} \mathcal{L}_m^{}\supset
-\frac{1}{2}m_\nu^{R}\bar{\nu}_L^{c}\nu_L^{}+\textrm{H.c.}\quad
\textrm{with}\quad\quad\quad\quad\quad\quad\quad\quad&&\nonumber\\
m_\nu^R=(y_l^{T}v_1^{}+\tilde{y}_l^{T}v_2^{})\frac{1}{f_R^\dagger
v_R^{}}M_R^\dagger \frac{1}{f_R^\ast
v_R^{}}(y_l^{}v_1^{}+\tilde{y}_l^{}v_2^{})\,,&&
\end{eqnarray}
which is the double seesaw \cite{mohapatra1986} for $M_R^{}\gg f
v_R^{}$ or the inverse seesaw \cite{mv1986} for $M_R^{}\ll f
v_R^{}$. In the double seesaw scenario, the right-handed neutrinos
should obtain a Majorana mass term much smaller than $M_R^{}$. For
$M_L^{}=M_R^{}$ (for the discrete left-right symmetry being a parity
transformation) or $M_L^{}=M_R^\dagger$ (for the discrete left-right
symmetry being a charge conjugation), the right-handed neutrinos
should also be much lighter than the left-handed triplets. In the
inverse seesaw scenario, the right-handed neutrinos and the neutral
components of the right-handed triplets should form the pseudo-Dirac
fermions with the masses of the order of $f_R^{} v_R^{}$. The
left-handed triplets and the charged components of the right-handed
triplets thus should be much lighter than the pseudo-Dirac fermions.

For an appropriate parameter choice, the canonical seesaw could
dominate over the double/inverse-type seesaw. For example, we can
obtain $ m_\nu^L\sim \mathcal{O}(0.1\,\textrm{eV})\gg m_\nu^R $ by
taking $v_L^{}\sim v_{1,2}^{}=\mathcal{O}(100\,\textrm{GeV})$,
$M_L^{}=M_R^{}(M_R^\dagger)=\mathcal{O}(\textrm{TeV})\ll
f_R^{}v_R^{}$, $f_L^{}=f_R^{}(f_R^\ast)\sim y_l^{}\sim
\tilde{y}_l^{}\sim \mathcal{O}(10^{-6}_{})$. It is possible to
detect the TeV-scale left-handed triplets at the LHC
\cite{abghhps2009}.

Our model can accommodate the leptogenesis \cite{fy1986} mechanism
to explain the matter-antimatter asymmetry in the universe. In the
double seesaw scenario, since the right-handed neutrinos are much
lighter than the neutral components of the left-handed triplets, the
leptogenesis should be realized by the decays of the right-handed
neutrinos \cite{fy1986}. In the inverse seesaw scenario, the neutral
components of the left-handed triplets are much lighter than the
pseudo Dirac fermions, which are composed of the right-handed
neutrinos and the neutral components of the right-handed triplets,
so that their decays should account for the leptogenesis
\cite{hlnps2003}. In the case of $M_R^{}\sim f v_R^{}$, the decays
of the left- and right-handed neutral fermions may both contribute
to the leptogenesis. Note that due to the trilinear couplings among
the Higgs bidoublet and the left- and right-handed Higgs doublets,
the left-handed Higgs doublet will mix to the Higgs doublets from
the Higgs bidoublet after the left-right symmetry breaking. So, if
the neutral components of the left-handed fermion triplets are
lighter than the left-handed Higgs doublet, they should decay into
the SM lepton and Higgs doublets as same as the right-handed
neutrinos. Not only the right-handed neutrinos but also the neutral
components of the left-handed triplets can mediate the loop
corrections in the decays of the right-handed neutrinos
\cite{bf2008}. The same story also exists in the decays of the
neutral components of the left-handed triplets.

Our model can be embedded into an $SO(10)$ GUT, where the Higgs
doublets, the Higgs bidoublet, the fermion doublets and the fermion
triplets belong to the following multiplets, respectively,
\begin{subequations}
\begin{eqnarray}
\textbf{10}_H^{}&=&\phi(\textbf{1},\textbf{2},\textbf{2},0)\oplus
(\textbf{3},\textbf{1},\textbf{1},-\frac{1}{3})\oplus
(\bar{\textbf{3}},\textbf{1},\textbf{1},\frac{1}{3})\,,\\
[1mm]
\textbf{16}_H^{}&=&\chi_L^\ast(\textbf{1},\textbf{2},\textbf{1},-1)\oplus
\chi_R^{}(\textbf{1},\textbf{1},\textbf{2},1)
\oplus (\textbf{3},\textbf{2},\textbf{1},\frac{1}{3})\nonumber\\
&& \oplus
(\bar{\textbf{3}},\textbf{1},\textbf{2},-\frac{1}{3})\,,\\
[1mm]
\textbf{16}_F^{}&=&l_L^{}(\textbf{1},\textbf{2},\textbf{1},-1)\oplus
l_R^c(\textbf{1},\textbf{1},\textbf{2},1)\oplus q_L^{}(\textbf{3},\textbf{2},\textbf{1},\frac{1}{3})\nonumber\\
&& \oplus
q_R^c(\bar{\textbf{3}},\textbf{1},\textbf{2},-\frac{1}{3})\,,\\
[1mm] \textbf{45}_F^{}&=&S(\textbf{1},\textbf{1},\textbf{1},0)\oplus
T_L^{}(\textbf{1},\textbf{3},\textbf{1},0)\oplus
T_R^c(\textbf{1},\textbf{1},\textbf{3},0)\nonumber\\
&&\oplus \Sigma(\textbf{3},\textbf{1},\textbf{1},\frac{4}{3}) \oplus
\Sigma^c_{}(\bar{\textbf{3}},\textbf{1},\textbf{1},-\frac{4}{3})\oplus
\Omega(\textbf{3},\textbf{2},\textbf{2},\frac{2}{3})\nonumber\\
&&\oplus
\Omega^c_{}(\bar{\textbf{3}},\textbf{2},\textbf{2},-\frac{2}{3})\oplus
(\textbf{8},\textbf{1},\textbf{1},0)\,.
\end{eqnarray}
\end{subequations}
At the left-right level, the terms involving the singlets $S$ should
include
\begin{eqnarray}
\mathcal{L}\supset -h_L^{}\bar{l}_L^{}\tilde{\chi}_L^{}S
-h_R^{}\bar{l}_R^{}\tilde{\chi}_R^{}S^c_{}
-\frac{1}{2}M_S^{}\bar{S}^c S+\textrm{H.c.}\,.
\end{eqnarray}
By performing the seesaw mechanism on
\begin{eqnarray}
\mathcal{L}_m^{}&\supset&-(y_l^{}v_1^{}+\tilde{y}_l^{}v_2^{})\bar{\nu}_L^{}\nu_R^{}
-h_L^{}v_L^{}\bar{\nu}_L^{}S-h_R^{}v_R^{}\bar{\nu}_R^{}S^c_{}\nonumber\\
&&-\frac{1}{2}M_S^{}\bar{S}S^c_{}+\textrm{H.c.}
\nonumber\\
&=&-\frac{1}{2}\left[\bar{\nu}_L^{}\,,\,\bar{\nu}_R^c\,,\,S^c_{}\right]\nonumber\\
[2mm] &&\times
\left[\begin{array}{ccc}0&y_l^{T}v_1^{}+\tilde{y}_l^{T}v_2^{}&h_L^{}v_L^{}\\
[2.mm]y_l^{} v_1^{}+\tilde{y}_l^{} v_2^{}
&0&h_R^\ast v_R^{}\\
[2.5mm]h_L^T v_L^{}&h_R^\dagger v_R^{} &
M_S^{}\end{array}\right]\left[\begin{array}{c}\nu_L^c\\
[2.mm]\nu_R^{}\\
 [2.mm] S\end{array}\right]\nonumber\\
 [2.mm]
 &&+\textrm{H.c.}\,,
\end{eqnarray}
we can read the contribution from the singlets $S$ to the neutrino
masses,
\begin{eqnarray}
\label{doubleinverse} \mathcal{L}_m^{}\supset
-\frac{1}{2}m_\nu^{S}\bar{\nu}_L^{}\nu_L^c+\textrm{H.c.}\quad
\textrm{with}\quad\quad\quad\quad\quad\quad\quad\quad&&\nonumber\\
m_\nu^S=(y_l^{}v_1^{}+\tilde{y}_l^{}v_2^{})\frac{1}{h_R^\dagger
v_R^{}}M_S^{} \frac{1}{h_R^\ast
v_R^{}}(y_l^{T}v_1^{}+\tilde{y}_l^{T}v_2^{})&&\nonumber\\
-[(y_l^{}v_1^{}+\tilde{y}_l^{}v_2^{})+
(y_l^{T}v_1^{}+\tilde{y}_l^{T}v_2^{})]\frac{v_L^{}}{v_R^{}}\,.\quad\quad~~&&
\end{eqnarray}
The first term is a double/inverse-type seesaw
\cite{mohapatra1986,mv1986} while the the second term is a linear
seesaw \cite{barr2003}. Therefore, we totally have two
double/inverse-type seesaw (from the singlets $S$ and the
right-handed triplets $T_R^{}$), a canonical seesaw (from the
left-handed triplets $T_L^{}$) and a linear seesaw (from the
singlets $S$). Only the double/inverse-type seesaw will survive
while the canonical and linear seesaw will disappear if we choose
the VEV $\langle\chi_L^{}\rangle$ to be zero as did in
\cite{cgr2010}. In the presence of a nonzero
$\langle\chi_L^{}\rangle$, the up- and down-type quarks can also
obtain the seesaw induced masses by integrating out the
$(\Sigma,\Sigma^c_{})$ and the $(\Omega,\Omega^c_{})$, respectively.
The VEV $\langle\chi_R^{}\rangle$ should be very large when it
accounts for the symmetry breaking of the $SU(2)_R^{}\otimes
U(1)_{B-L}^{}$ down to the $U(1)_Y^{}$. Accordingly, the
$\textbf{45}_F^{}$ should be very heavy to guarantee the quark
seesaw. Actually, the $\textbf{45}_F^{}$ could not be very light if
they have the Yukawa couplings with the Higgs multiplets (such as a
$\textbf{210}_H^{}$) which are responsible for breaking the
$SO(10)$. In case the $\textbf{45}_F^{}$ are not very heavy, we may
introduce a new $\textbf{45}_H^{}=
(\textbf{1},\textbf{3},\textbf{1},0)\oplus
(\textbf{1},\textbf{1},\textbf{3},0)\oplus ... $ to break the
$SU(2)_R^{}$ down to a $U(1)_R^{}$ \cite{mrv2005}. Then the
$U(1)_R^{}\otimes U(1)_{B-L}^{}$ is broken down to the $U(1)_Y^{}$
by the $\textbf{16}_H^{}$. We hence can take the
$\langle\chi_R^{}\rangle$ at a lower scale such as the TeV.

In this paper we studied the left-right symmetric model with the
$SU(2)$-triplet fermions. The neutral components of the
$SU(2)_L^{}$-triplet fermions lead to a canonical seesaw while the
neutral components of the $SU(2)_R^{}$-triplet fermions associated
with the right-handed neutrinos result in a double/inverse-type
seesaw. The neutrinos thus can naturally obtain the small masses.
Meanwhile, the leptogenesis can work through the decays of the left-
and/or right-handed neutral fermions. We further indicated that our
model could be realized in an $SO(10)$ GUT with the fermionic
$\textbf{45}$ multiplets.

\textbf{Acknowledgement}: This work is supported by the Alexander
von Humboldt Foundation.


\begin{thebibliography}{99}



\bibitem{ps1974}
J.C. Pati and A. Salam, Phys. Rev. D \textbf{10}, 275 (1974); R.N.
Mohapatra and J.C. Pati, {\it ibid.} \textbf{11}, 566 (1975); R.N.
Mohapatra and J.C. Pati, {\it ibid.} \textbf{11}, 2558 (1975); R.N.
Mohapatra and G. Senjanovi\'{c}, {\it ibid.} \textbf{12}, 1502
(1975).





\bibitem{berezhiani1983}
Z.G. Berezhiani, Phys. Lett. B \textbf{129}, 99 (1983); D. Chang and
R.N. Mohapatra, Phys. Rev. Lett. \textbf{58}, 1600 (1987); S.
Rajpoot, Phys. Lett. B \textbf{191}, 122 (1987); A. Davidson and
K.C. Wali, Phys. Rev. Lett. \textbf{59}, 393 (1987).



\bibitem{minkowski1977}
P. Minkowski, Phys. Lett. B \textbf{67}, 421 (1977); T. Yanagida, in
{\it Proc. of the Workshop on Unified Theory and the Baryon Number
of the Universe}, ed. O. Sawada and A. Sugamoto (KEK, Tsukuba,
1979), p. 95; M. Gell-Mann, P. Ramond, and R. Slansky, in {\it
Supergravity}, ed. F. van Nieuwenhuizen and D. Freedman (North
Holland, Amsterdam, 1979), p. 315; S.L. Glashow, in {\it Quarks and
Leptons}, ed. M. L\'{e}vy {\it et al.} (Plenum, New York, 1980), p.
707; R.N. Mohapatra and G. Senjanovi\'{c}, Phys. Rev. Lett.
\textbf{44}, 912 (1980).


\bibitem{mw1980}
M. Magg and C. Wetterich, Phys. Lett. B \textbf{94}, 61 (1980); J.
Schechter and J.W.F. Valle, Phys. Rev. D \textbf{22}, 2227 (1980);
T.P. Cheng and L.F. Li, Phys. Rev. D \textbf{22}, 2860 (1980); G.
Lazarides, Q. Shafi, and C. Wetterich, Nucl. Phys. B \textbf{181},
287 (1981); R.N. Mohapatra and G. Senjanovi\'{c}, Phys. Rev. D
\textbf{23}, 165 (1981).


\bibitem{barr2003}
S.M. Barr, Phys. Rev. Lett. \textbf{92}, 101601 (2004).

\bibitem{mohapatra1986}
R.N. Mohapatra, Phys. Rev. Lett. \textbf{56}, 561 (1986).

\bibitem{mv1986}
R.N. Mohapatra and J.W.F. Valle, Phys. Rev. D \textbf{34}, 1642
(1986).



\bibitem{perez2009}
P.F. Perez, JHEP \textbf{0903}, 142 (2009).

\bibitem{flhj1989}
R. Foot, H. Lew, X.G. He, and G.C. Joshi, Z. Phys. C \textbf{44},
441 (1989).

\bibitem{abghhps2009}
A. Arhrib {\it et al.}, Phys. Rev. D \textbf{82}, 053004 (2010).

\bibitem{fy1986}
M. Fukugita and T. Yanagida, Phys. Lett. B \textbf{174}, 45 (1986).

\bibitem{hlnps2003}
T. Hambye, Y. Lin, A. Notari, M. Papucci, and A. Strumia, Nucl.
Phys. B \textbf{695}, 169 (2004).


\bibitem{bf2008}
S. Blanchet and P.F. Perez, JCAP \textbf{0808}, 037 (2008).


\bibitem{cgr2010}
J. Chakrabortty, S. Goswami, and A. Raychaudhuri, Phys. Lett. B
\textbf{698}, 265 (2011).

\bibitem{mrv2005}
M. Malinsky, J.C. Rom\~{a}o, and J.W.F. Valle, Phys. Rev. Lett.
\textbf{95}, 161801 (2005).

\end{thebibliography}
\end{document}